# PPI-Net connects molecular protein interactions to functional processes in disease

Kyle Higgins[1], Guadalupe Gonzalez[2,3,4], Dennis Veselkov[1,4], Ivan Laponogov[1], Kirill Veselkov[1*]

[1]Division of Cancer, Department of Surgery and Cancer, Faculty of Medicine, Imperial College London, London, UK.
[2]F. Hoffmann-La Roche Ltd, Basel, Switzerland.
[3]Prescient Design, Genentech, South San Francisco, CA, USA.
[4]Department of Computing, Imperial College London, U.K.
[*]To whom correspondence should be addressed: kirill.veselkov04@imperial.ac.uk

**Abstract**

Understanding how molecular alterations propagate across biological systems to drive disease remains a central challenge. Although high-throughput profiling enables comprehensive characterization of tumor states, most models neglect structured biological relationships or lack interpretability across scales. Here we present PPI-Net, a hierarchical graph neural network that integrates protein–protein interaction (PPI) networks with pathway-level representations to model disease from molecular interactions to functional processes. Patient-specific molecular profiles are embedded within a shared interaction network from STRING and propagated through a multi-layer Reactome hierarchy using graph attention, enabling aggregation of gene-level signals into higher-order biological programs. Across RNA-seq data from ten cancer types from The Cancer Genome Atlas, PPI-Net achieves robust predictive performance, with balanced accuracy exceeding 90% in multiple cohorts. Comparative analysis on RNA-Seq data from breast cancer demonstrated that PPI-Net's integration of the Reactome hierarchy improved balanced accuracy by 6.7% relative to a PPI-only model, while hierarchical multi-level supervision improved balanced accuracy by 12.3% relative to using only a single top-level prediction head. Applying a multi-omics approach using RNA-seq and methylation data improves model interpretation, recovering canonical oncogenic modules, including TP53–AKT signaling and stress-response pathways, while revealing convergence onto coherent programs such as ion signaling and cellular responses to stimuli. These results demonstrate that integrating interaction networks with pathway hierarchies enables accurate prediction while providing mechanistic insight into cancer biology.

## Introduction

Cancer is a multi-scale disease[1] arising from the accumulation of molecular alterations that propagate across biological systems to drive dysregulated cellular behavior.[2,3] At the molecular level, genetic and epigenetic perturbations alter gene expression and protein function.[4] However, these changes rarely act in isolation; instead, they perturb interconnected protein–protein interaction (PPI) networks[5-7], which in turn influence higher-order biological processes such as signaling, metabolism and stress response. Understanding how localized molecular changes give rise to coordinated functional programs remains a central challenge in cancer biology.

Breast cancer provides a particularly well-characterized context in which to study the relationship between molecular alterations and system-level biological processes. Extensive genomic and epigenomic profiling[8] has identified key driver pathways, including TP53 signaling[9], PI3K–AKT activation[10] and dysregulated proteostasis[11,12], alongside emerging roles for ion channel activity[13-15] and metabolic adaptation[16,17]. The availability of large, well-annotated cohorts has further enabled systematic evaluation of computational models linking molecular features to functional programs.[18]

High-throughput technologies such as RNA sequencing[19] and DNA methylation profiling[20] have enabled comprehensive characterization of tumor molecular states across large patient cohorts. Resources such as The Cancer Genome Atlas (TCGA)[18] provide unprecedented opportunities to link molecular features to clinical phenotypes across many cancers. However, many computational approaches treat these data as independent features, neglecting the structured biological relationships that govern cellular function. As a result, models may achieve strong predictive performance but lack interpretability and mechanistic insight.

Network-based approaches offer a principled framework for integrating molecular data with prior biological knowledge. PPI networks capture the physical and functional dependencies between proteins, enabling the identification of dysregulated interaction modules in disease.[6] Complementarily, curated pathway databases such as Reactome[21] organize biological processes into hierarchical structures, reflecting the organization of cellular systems from molecular events to organism-level functions. Several methods have leveraged either PPI networks or pathway annotations to improve predictive modelling and interpretability[22-25], with multi-omics capable of improving results.[26,27] Notably, the P-NET[28] framework introduced a hierarchical architecture in which molecular features are propagated through gene and pathway layers constrained by curated biological relationships from Reactome, enabling a highly interpretable pathway level image. However, PPI and functional pathway approaches are typically applied in isolation, limiting their ability to capture the full cascade from molecular perturbation to system-level dysfunction.

Graph neural networks (GNNs) have recently emerged as powerful tools for modelling structured biological data, enabling the integration of topology and node-level features within a unified framework. In cancer research, GNNs have been applied to molecular interaction networks to improve prediction tasks such as tumor classification[29-31] and drug response[32,33]. Nevertheless, most existing architectures operate on a single level of biological organization, either focusing on molecular interaction graphs or pathway-level abstractions, without explicitly linking the two.

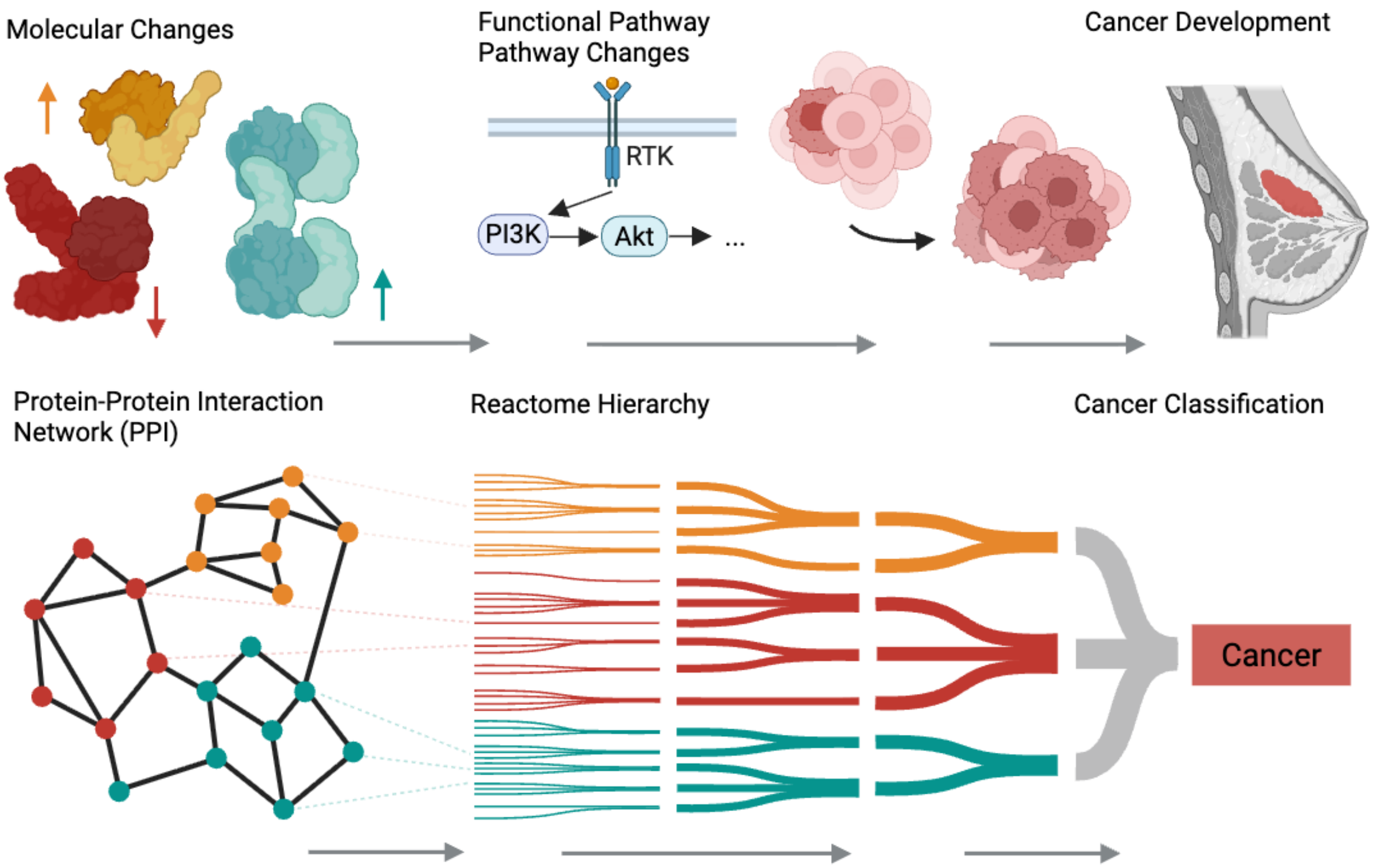

**Figure 1 Model architecture linking molecular alterations to pathway-level representations for cancer classification.**
Schematic overview of the modelling framework. Molecular changes at the protein level are first represented within a protein–protein interaction (PPI) network, capturing local interaction structure. Signals are then propagated through a hierarchical organization of Reactome pathways, aggregating gene-level information into progressively higher-order functional programs. These pathway-level representations reflect coordinated biological processes, including signaling, stress response and metabolism, and are integrated to produce a final cancer classification output. Arrows indicate the direction of information flow from molecular features to system-level prediction. Created in BioRender. Higgins, K. (2026) https://BioRender.com/cwmy22c

Here, we introduce PPI-Net, a hierarchical graph neural network framework that bridges molecular interaction networks and pathway-level representations to model disease processes across biological scales. (**Figure 1**) In contrast to prior approaches that operate directly on gene-level features, PPI-Net first models patient-specific molecular states within a protein interaction network, enabling the capture of local molecular context before propagating signals through a structured pathway hierarchy using a sequential series of bipartite graph attention layers, allowing the model to aggregate signals from genes to pathways and ultimately to system-level biological programs. This "conveyor-belt" design provides a graph-native mechanism for hierarchical biological aggregation while preserving level-specific interpretability.

A key feature of PPI-Net is its ability to provide interpretable, multi-level attributions that connect model predictions to biologically meaningful mechanisms. By combining gradient-based importance with attention-driven information flow across the pathway hierarchy, the model identifies coherent modules spanning protein interactions, genes and pathways. This enables the recovery of known oncogenic processes while also highlighting less well-characterized biological signals.

We apply PPI-Net across multiple cancer types and omics layers using TCGA datasets and demonstrate robust predictive performance for tumor classification. Importantly, the model enables interpretation of predictions in terms of coherent biological programs spanning molecular interactions to pathway-level processes. Together, these results illustrate how integrating molecular interaction networks with hierarchical pathway knowledge can provide both accurate predictions and mechanistic insight into cancer biology.

## Methods

### *Patient graph construction*

Patient-level molecular data, including RNA-seq and methylation, was obtained from The Cancer Genome Atlas (TCGA). We filtered cancer types to require at least 20 samples per minority class for RNA-seq, resulting in a dataset spanning multiple tumor types including breast (BRCA), colorectal (COAD), head and neck (HNSC), liver (LIHC), lung adenocarcinoma (LUAD), lung squamous (LUSC), prostate (PRAD), gastric (STAD), thyroid (THCA), and endometrial (UCEC) cancers. For each patient, gene level features derived from RNA sequencing were integrated by log normalization (and DNA methylation, for the RNA and average beta value per gene methylation BRCA alternative model) and mapped onto the STRING protein-protein interaction (PPI) network.

Each patient was represented as a graph $G = (V, E)$, where nodes correspond to proteins and edges represent experimentally supported physical interactions (top 100,000 highest confidence score). The graph topology was fixed across all samples, while node features were patient-specific omics, enabling the model to learn consistent structural representations across individuals. Protein identifiers were mapped from Ensmbl protein IDs to HGNC gene symbols to enable downstream integration with pathway-level annotations from Reactome.

### *Construction of the Reactome pathway hierarchy*

The construction of the Reactome pathway hierarchy was adapted in part from previously described implementations in biologically informed neural network models, including P-NET[28]. Specifically, we utilized curated pathway relationship files from Reactome and hierarchical organization strategies consistent with P-NET[28], while extending the representation to enforce compatibility with graph-based message passing. Only homo sapiens pathways were retained in the analysis.

A directed acyclic graph was assembled in which nodes represent pathways organized into multiple hierarchical levels, with a synthetic root node connected to all top-level pathways. Terminal pathways were linked to constituent genes, forming a multi-layer structure spanning pathways-level abstractions down to molecular entities.

To ensure a consistent depth across all branches of the hierarchy, terminal nodes were extended using auxiliary edges such that all paths from the root to gene nodes had equal length. This resulted in a balanced hierarchical graph comprising five pathways layers and a terminal gene layer, enabling uniform message passing across levels.

*Model architecture*

The model consists of two sequential graph neural network modules that bridge molecular interactions and biological pathways. A mixture of Pytorch Geometric modules and custom code were used to form this architecture. In the first stage, patient-specific PPI graphs were processed using a graph convolutional network (GCN). Multiple GraphConv layers were applied to propagate information across the interaction network, followed by non-linear activation and dropout regularization. This module produces latent node embeddings that capture the local molecular context of each protein.

These node embeddings were then mapped onto the Reactome gene layer. Because multiple proteins may correspond to a single gene, mappings were resolved using a deterministic assignment, and genes without corresponding PPI nodes were initialized with zero embeddings. This produces a gene-level representation aligned with the Reactome hierarchy.

In the second stage, a hierarchical graph attention network (GAT) was applied over the Reactome structure. Information was propagated in a bottom-up manner, beginning at the gene layer and proceeding through successive pathway levels towards the root. At each layer transition, attention-based message passing was performed using bipartite graph attention convolutions, enabling the model to learn weighted contributions of lower-level nodes to higher-level pathway representations. Residual connections and layer normalization were applied at each layer to stabilize training and preserve information across hierarchical steps. At each layer, a decision head generated a scalar prediction logit, reflecting the contribution of that level to the overall prediction.

*Hierarchical prediction and loss function*

Predictions from all hierarchical levels were aggregated using a weighted combination of layer-specific logits. Specifically, logits were combined using a geometrically decaying weighting scheme controlled by a parameter $\alpha$, such that deeper layers contribute progressively less to the final prediction. This formulation allows the model to integrate information across multiple biological scaled while prioritizing higher-level abstractions.

The final prediction was obtained by applying a sigmoid transformation to the weighted logit. Model parameters were optimized using weighted binary cross-entropy with logits, with class imbalance addressed through a positive class weighting proportional to the ratio of negative to positive samples.

*Training and evaluation*

For each cancer type, data were split into training, validation, and tests sets (20% held for testing, with remaining 80% split into training and validation using another 80:20 split; 64:16:20 total) using stratified sampling to preserve class balance. Models were trained using the Adam optimizer with weight decay regularization, and early stopping was applied based on validation balanced accuracy. Performance on test set was evaluated using balanced accuracy.

| Cancer | Cancer Type | Mean BA | BA SD | N samples | Class Ratio Cancer:Normal |
|---|---|---|---|---|---|
| **TCGA-BRCA** | Breast | **0.91** | 0.07 | 1224 | 9.8:1 |
| **TCGA-COAD** | Colon / Colorectal | **0.84** | 0.16 | 522 | 11.7:1 |
| **TCGA-HNSC** | Head & Neck | **0.83** | 0.10 | 564 | 11.8:1 |
| **TCGA-LIHC** | Liver (Hepatocellular) | **0.92** | 0.06 | 421 | 7.4:1 |
| **TCGA-LUAD** | Lung (Adenocarcinoma) | **0.93** | 0.03 | 599 | 9.2:1 |
| **TCGA-LUSC** | Lung (Squamous) | **0.96** | 0.03 | 562 | 10.0:1 |
| **TCGA-PRAD** | Prostate | **0.75** | 0.07 | 553 | 9.6:1 |
| **TCGA-STAD** | Stomach / Gastric | **0.89** | 0.07 | 448 | 11.4:1 |
| **TCGA-THCA** | Thyroid | **0.81** | 0.08 | 564 | 8.6:1 |
| **TCGA-UCEC** | Uterine (Endometrial) | **0.76** | 0.17 | 588 | 15.8:1 |

**Table 1 Predictive performance of PPI-Net across cancer types.**
Balanced accuracy (BA) for cancer classification using PPI-Net across ten tumor types from The Cancer Genome Atlas (TCGA). Values represent the mean balanced accuracy and standard deviation (s.d.) across five cross-validation folds. Performance exceeds 0.9 in multiple cohorts, including lung (LUAD, LUSC), liver (LIHC) and breast (BRCA) cancers, indicating robust generalization across diverse tumor contexts. Lower performance in selected cancers (for example, PRAD and UCEC) may reflect differences in cohort size and molecular heterogeneity.

*Explainability and attribution framework*

To interpret the model predictions, we developed a multi-level attribution framework that quantified the contributions of molecular interactions, genes, and pathways. At the PPI level, node importance scores were computed using a gradient-based attribution method applied to learned node embeddings. Edge importance was subsequently derived from node scores, allowing identification of high-impact interaction subgraphs. For each patient graph, the top 30 attributed PPI nodes and top 100 edges among these nodes were retained, and edge scores were then averaged across cancer samples.

At the gene level, importance was quantified using gradient magnitudes with respect to gene embeddings in the Reactome layer. These scores were aggregated across samples to identify consistently influential genes. At the pathway level, we defined a directed flow metric that combines attention weights with gradient-based importance. Specifically, for each edge in the Reactome hierarchy, flow was computed as the product of the attention coefficient and the importance of the destination node. This formulation captures both the strength of information propagation and the sensitivity of downstream nodes, enabling a biologically interpretable measure of signal transmission across the hierarchy.

| Model / Ablation | Mean BA | SD (BA) | Mean Acc | SD (Acc) | Mean F1 | SD (F1) |
|---|---|---|---|---|---|---|
| **Full (PPI + Reactome + multi-level loss)** | **0.93** | 0.04 | 0.93 | 0.04 | 0.96 | 0.03 |
| **PPI-only** | **0.87** | 0.02 | 0.87 | 0.06 | 0.92 | 0.04 |
| **Top-layer-only loss** | **0.83** | 0.19 | 0.90 | 0.04 | 0.94 | 0.02 |

**Table 2 Ablation analysis of PPI-Net architecture on the TCGA-BRCA cohort using five-fold cross-validation.** Comparison of the full PPI-Net architecture against two ablated variants: (i) a PPI-only model without Reactome hierarchical pathway propagation, and (ii) a model retaining the Reactome hierarchy but trained using only the top-layer prediction loss, removing hierarchical multi-level supervision. Metrics are reported as mean ± standard deviation across folds for balanced accuracy (BA), overall accuracy (Acc), and F1 score. The full model achieved the highest balanced accuracy and lowest performance variability, demonstrating that both Reactome-guided hierarchical aggregation and multi-level supervision contribute substantially to predictive performance and model stability.

Aggregated edge flows were used to construct Sankey-style representations of hierarchical information flow. High-level biological programs were identified by selecting top pathways based on incoming flow and recursively tracing their upstream contributors, yielding interpretable multi-scale representation of disease-associated processes.

*Hyperparameter Tuning*

Hyperparameters were selected using a structured one-factor-at-a-time optimization strategy rather than exhaustive grid search. This approach was designed to identify a strong configuration efficiently while avoiding the combinatorial redundancy of full hyperparameter enumeration. Optimization was performed using the BRCA cohort and RNA-seq-derived graph inputs. Candidate values were evaluated by training the PPI–Reactome (PPI-Net) model on a stratified training split and monitoring balanced accuracy on the validation set. Early stopping was applied during each sweep to limit unnecessary training once validation performance plateaued.

The optimization procedure began from a fixed backbone architecture consisting of a three-layer PPI graph encoder with hidden dimension 8, output dimension 8, ReLU activation, no feature dropout in the PPI encoder, and no dropout across Reactome hierarchy layers. During optimization, each candidate parameter was varied while all other settings were held constant. The hyperparameter search space comprised the following ranges: number of PPI graph convolution layers {2, 3, 4}; hidden dimension {8, 16, 32, 64}; PPI feature dropout {0.0, 0.1, 0.3}; number of graph attention heads {1, 2, 4}; attention dropout {0.0, 0.1}; Reactome-layer dropout schedules {(0,0,0,0,0,0),(0.2,0.2,0.2,0.2,0.2,0.2)}; learning rate $\{10^{-2}, 5\times10^{-3}, 10^{-3}, 5\times10^{-4}, 10^{-4}, 5\times10^{-5}, 10^{-5}\}$; weight decay $\{0, 10^{-4}\}$; and hierarchical loss-weighting coefficient $\alpha \in \{0.5, 0.7, 0.9\}$.

This tuning procedure selected an architecture with three PPI graph convolution layers, hidden and output dimension of 8, two graph attention heads, zero attention dropout, zero weight decay, learning rate $=10^{-3}$, and hierarchical loss-weighting coefficient $\alpha$=0.7. These parameters were subsequently fixed for final model training and interpretation experiments.

*Ablation Analysis*

To assess the contribution of the Reactome hierarchy and hierarchical multi-level supervision, we performed a five-fold cross-validation ablation study using the TCGA-BRCA cohort. Three model variants were compared: the full PPI-Net model, consisting of a PPI graph encoder followed by Reactome-based hierarchical graph attention with multi-level weighted loss; a PPI-only model, in which the Reactome hierarchy was removed and graph-level PPI embeddings were used directly for classification; and a top-layer-only loss model, which retained the full PPI-to-Reactome architecture but optimized only the highest-level decision head, removing hierarchical supervision from intermediate pathway layers. For each fold, data were split using the stratified five-fold cross-validation and training procedure described above. Test performance was evaluated on the held-out fold using balanced accuracy, accuracy, and F1 score. The same PPI encoder dimensionality, Reactome architecture, attention-head configuration, batch size, and optimization settings were used across comparable ablations to isolate the effect of the removed component.

## Results

*Robust predictive performance across cancer types*

The PPI-Net model demonstrated a consistently high predictive performance for cancer prediction using RNA-seq projected to STRING across a diverse set of cancer types. (**Table 1**) Mean balanced accuracy in the test dataset exceeded 90% in multiple cohorts, including lung adenocarcinoma (LUAD: 93.4%), lung squamous carcinoma (LUSC: 96.2%), liver (LIHC: 91.7%), and breast (BRCA: 91.4%) cancers, indicating strong generalization across distinct tumor contexts. Moderate performance was observed in other cancers, such as prostate (PRAD: 75.0%) and endometrial (UCEC: 76.4%), likely due to factors including fewer samples and tumor heterogeneity. To determine the effect of including multiple omics inputs into the method, both RNA-seq and methylation (average beta score per gene) were used in an alternative model, achieving 98.0% BA on validation and 88.3% BA on test set. These results indicate that integrating molecular interaction networks with hierarchical biological networks provides a robust framework for patient prediction across cancer types.

*Ablation study demonstrates contributions of Reactome hierarchy and multi-level supervision*

Ablation analysis showed that both the Reactome hierarchy and multi-level decision-head supervision contributed substantially to predictive performance. The full PPI-Net model achieved the highest mean balanced accuracy across five folds, with a mean balanced accuracy of 0.93 ± 0.04, mean accuracy of 0.93 ± 0.04, and mean F1 score of 0.96 ± 0.03. (**Table 2**)

Removing the Reactome hierarchy reduced performance to a mean balanced accuracy of 0.87 ± 0.02. Thus, incorporating hierarchical Reactome-based pathway aggregation improved balanced accuracy by 5.9 percentage points over the PPI-only model,  corresponding to a 6.7% relative

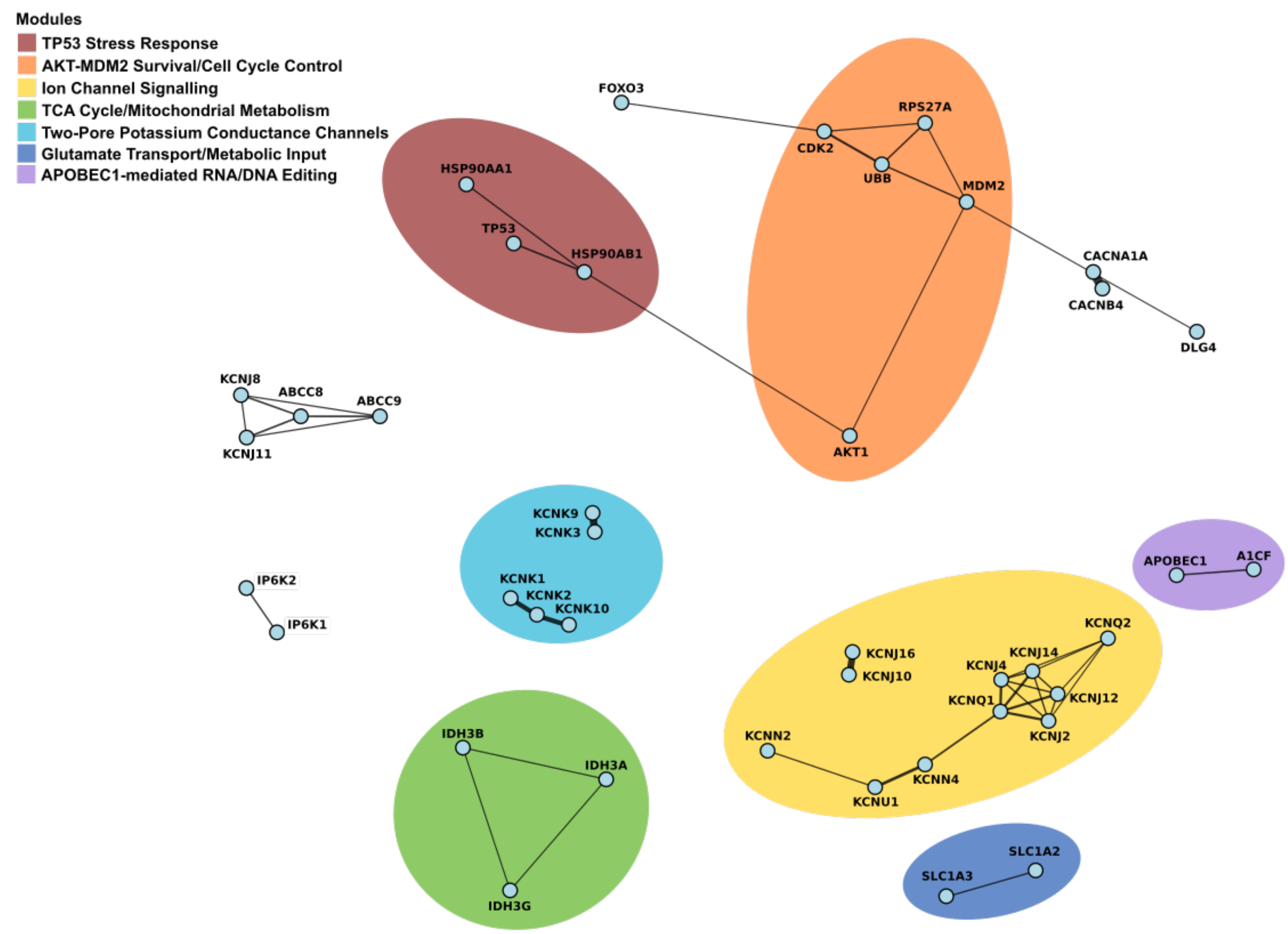

**Figure 2 Protein–protein interaction modules associated with model-derived gene importance in breast cancer.**

Subnetwork of the protein–protein interaction graph showing genes with high attribution. Nodes represent proteins and edges denote known physical interactions. Genes are organized into modules based on connectivity and functional coherence, corresponding to ion channel signaling (yellow; KCN family), AKT–MDM2/cell-cycle signaling (orange), stress and proteostasis (red; TP53–HSP90 axis), mitochondrial metabolism (green; IDH3 complex), potassium conductance channels (cyan; KCNK family), glutamate/synaptic transport (blue; SLC1A family), and RNA editing (purple; APOBEC1–A1CF). Node layout is optimized for visualization and does not reflect spatial or quantitative relationships.

improvement. This indicates that pathway-level organization provides predictive information beyond molecular interaction structure alone.

The top-layer-only loss model, which preserved the PPI-to-Reactome architecture but removed intermediate hierarchical supervision, achieved a lower mean balanced accuracy of 0.83 ± 0.19. Compared with this model, the full multi-level supervision strategy improved balanced accuracy by 10.2 percentage points, corresponding to a 12.3% relative improvement. The larger variance observed for the top-layer-only model suggests that relying exclusively on the final pathway-level decision head reduces training stability and generalization.

*Model attribution recovers canonical oncogenic interaction modules in multi-omics context*

RNA-only models produced coherent but less disease-specific attribution patterns, capturing general cellular processes such as ion transport, adhesion, and signaling. In contrast, integration of DNA methylation sharpened these representations toward canonical breast cancer pathways, including TP53-mediated stress responses, AKT–MDM2 signaling, and proteostasis networks. (**Figure 2**) This suggests that epigenetic information helps contextualize transcriptional signals within disease-relevant regulatory programs. Interpretation of the learned PPI representations revealed that the model consistently prioritized well-established oncogenic interaction modules. Prominent among these was the TP53–MDM2–AKT1 regulatory axis, a central hub governing apoptosis, DNA damage response and cell proliferation. The emergence of this module across samples indicates that the model captures fundamental regulatory mechanisms underlying tumor behavior.

In addition, strong importance was assigned to molecular chaperones HSP90AA1 and HSP90AB1, consistent with their established role in maintaining proteostasis in cancer cells and their status as therapeutic targets. Components of the ubiquitin–proteasome system, including RPS27A and UBB, were also highly ranked, reflecting the importance of protein turnover and stress adaptation in tumor progression.

Notably, the model highlighted clusters of ion channel proteins, including members of the KCNQ, KCNJ and KCNK families, as well as calcium channel subunits such as CACNA1A and CACNB4. These findings are consistent with emerging evidence that ion channel dysregulation contributes to cancer cell survival and proliferation, suggesting that the model captures both canonical and less well-characterized biological signals.

*Hierarchical pathway attribution reveals biologically coherent programs*

Projection of model predictions onto the Reactome hierarchy revealed structured patterns of information flow that correspond to coherent biological programs. (**Figure 3**) At the highest levels of the hierarchy, pathways related to cellular responses to external stimuli emerged as dominant contributors, integrating signals from stress response, DNA damage signaling and protein folding pathways. Key contributors included heat shock protein 90 components HSP90AA1 and HSP90AB1, which are associated with decreased breast cancer survival[34] and a popular therapeutic choice due to its chaperoning of the estrogen receptor, TP53, receptor tyrosine kinases (erbB family), and AKT.[35-37] Our model emphasizes their chaperone cycle for steroid hormone receptors as a cellular response to stress. Both UBB and RSP27A are emphasized in the downstream pathway level by contributions to cellular responses to stress and to external stimuli, hallmarks of cancer progression. Our pathway analysis emphasizes the importance of UBB interacting with HSF1, a known therapeutic target in breast cancer[38] whose elevated presence in the nucleus is associated with poor prognostic outcomes.[39] UBB has an established role in breast cancer, with knockdown experiments demonstrating a lower proliferation rate of cancer cells.[40] It may also control HSP90 clients in HER-2 overexpressing subtypes, tying together these components.[41] RPS27A has yet to be strongly characterized as causative in breast cancer, however it has recently been highlighted its regulatory role in lung adenocarcinoma[42] as a candidate for pan-cancer prognosis and immunotherapy.[43]

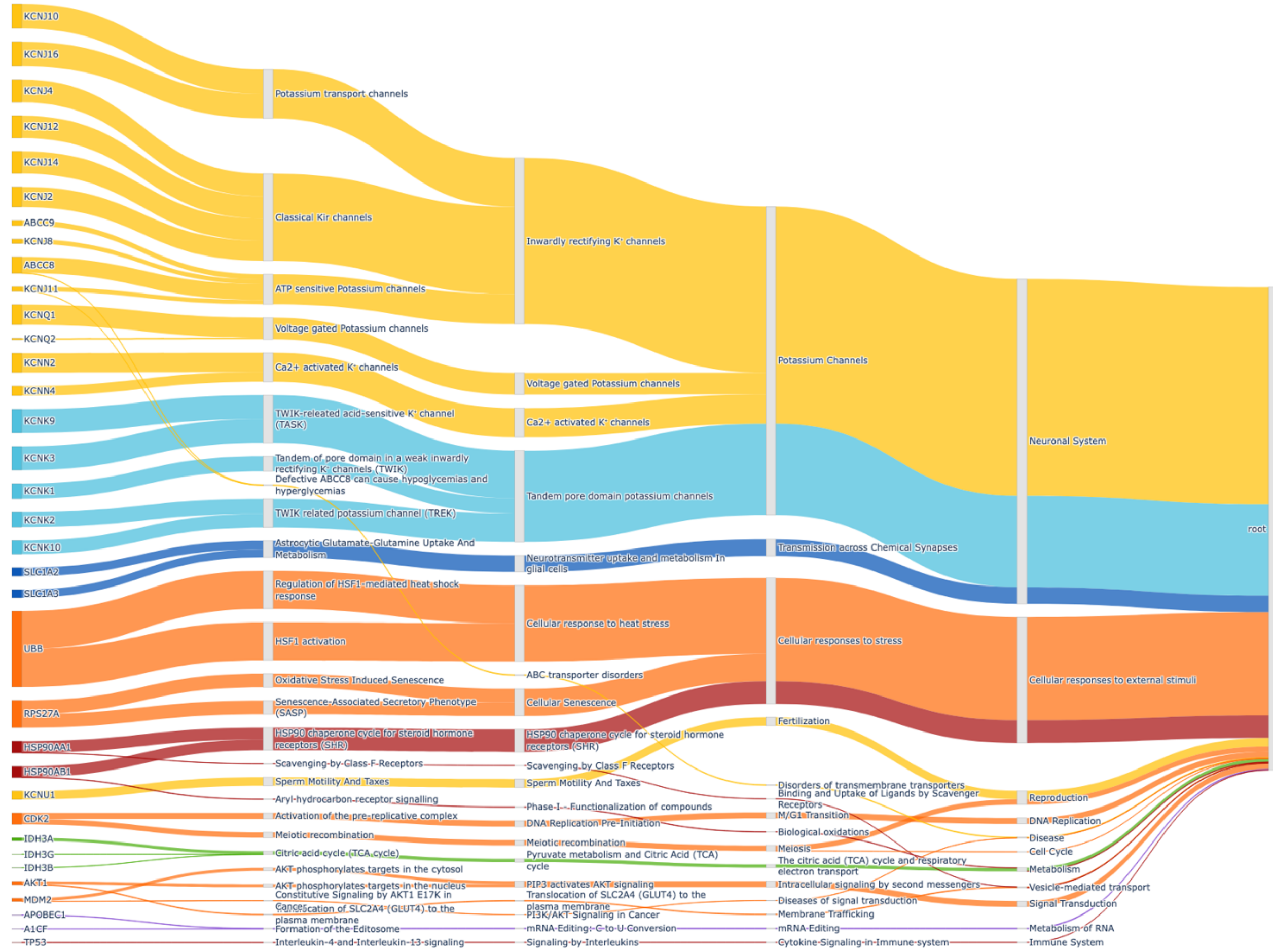

**Figure 3 Gene-to-pathway attribution reveals convergent ion-channel and stress-response programs in breast cancer.**

Sankey diagram showing model-derived attribution flow from genes to higher-order biological processes through the Reactome pathway hierarchy. Individual genes (left) are connected to progressively higher levels of pathway organization (L4–L1), culminating in a root node representing the integrated model output. Link widths are proportional to the relative contribution of each path, highlighting dominant routes of information flow. Distinct colored streams correspond to functional modules defined in the protein–protein interaction network (Fig. 1), including ion channel signaling (yellow), stress and proteostasis (red), AKT–MDM2/cell-cycle signaling (orange), mitochondrial metabolism (green), two-pore potassium conductance channels (cyan), glutamate/synaptic transport (blue) and RNA editing (purple). The model identifies strong convergence of potassium channel genes onto neuronal system pathways and cellular responses to external stimuli, alongside parallel contributions from stress-response and signaling pathways, indicating coordinated multi-pathway drivers of the predicted cancer state.

TP53, MDM2, and AKT1 and the multi-scale effects of their downstream pathways are emphasized by our model. Dysregulation of the tumor suppressor TP53 is a hallmark of many human cancers[44], and is the most frequently mutated gene in breast cancer (30% of breast cancers).[45] It drives faster tumor growth (tumor suppression loss), chemoresistance, and more aggressive phenotype.[46,47] These pathways are emphasized by our Reactome pathway analysis, with AKT1 and MDM2 both linking to AKT-driven phosphorylation, and downstream PIP3 signaling, cellular signaling, and others. MDM2 and MDM3 are known oncoproteins that inhibit p53, promoting cancer cell survival.[48] AKT1 is also a hallmark for breast cancer, promoting the suppression of TP53 through phosphorylation-dependent activation of MDM2, with its overactivation acting as a major driver of progression and resistance to therapy.[49] These axes link the TP53, AKT and MDM2 systems identified at the PPI level, providing a multi-scale view of tumor adaptation to TP53 suppression.

A second prominent program involved neuronal system pathways, driven by contributions from potassium and calcium channel genes. This included inwardly rectifying (KCNJ), two-pore domain (KCNK), voltage-gated (KCNQ), and calcium-activated potassium channels. Several of these members have identified roles in breast cancer biology, including the KNCK family (e.g., KCNK1, KCNK3, KCNK9, KCNK10), which were enriched in multiple pathways, joining TWIK-related acid-sensitive $K^+$ channels (TASK) to TWIK related potassium channels (TREK) in the broader process of tandem pore domain potassium channels. Prior analyses have noted differential expression of these genes in breast cancer, with poorer overall survival in the presence of dysregulated expression[50], and KCNK1 promoting increased proliferation and invasion of breast cancer cells in vivo and in vitro[51] and KCNK9 overexpression promoting tumor formation and hypoxia resistance[52].

Inward rectifier channel (KCNJ11 and KCNJ3) and calcium channel (KCNN2 and KCNN4) signaling are emphasized by our model. KCNJ11[53] and KCNJ3[54] have been reported as an prognostic biomarkers for breast cancer (ER-positive, in the case of KCNJ3). Our model further links calcium channel signaling (KCNN2 and KCNN4) to potassium-related signaling as a downstream process, emphasizing an interaction between signaling types. Calcium channels are increasingly recognized as regulators of proliferation, apoptosis , migration, and metabolic adaptation in breast cancer, via control of membrane potential and ion flux.[55-57] The high emphasis our model puts on this suggests that ion-channel-mediated signaling may play a broader role in tumor biology, potentially through regulation of membrane potential and intracellular signaling dynamics.

Metabolic pathways associated with digestion and absorption also exhibited strong contributions, indicating a role for nutrient acquisition and metabolic adaptation. These signals were closely linked to proteasomal degradation pathways, suggesting coordination between nutrient scavenging and protein turnover.[58] Ubiquitin-proteasome interactions are also a known component to breast cancer, tying into the observation of UBB importance above.[59]

Additional programs related to reproductive processes and vesicle-mediated transport were also identified. The activation of reproduction-related pathways is consistent with dedifferentiation and the reactivation of developmental programs in cancer,[60] while vesicle-mediated transport pathways may reflect increased membrane trafficking and resource acquisition in tumor cells.

## Discussion

The design of PPI-Net was conceptually inspired by an interest in expanding biologically informed neural network architectures such as omics-driven PPI graph classification[29,31] to the multi-scale domain which drives disease progression. We achieved this by adapting the pathway structure found in P-NET[28] to the graph AI domain, which demonstrated the value of embedding pathway hierarchies into predictive models. By incorporating protein–protein interaction structure upstream of pathway aggregation, PPI-Net extends traditional PPI paradigms to capture multi-scale biological organization from molecular interactions to functional processes.

A central methodological contribution of PPI-Net is therefore the conversion of a curated pathway hierarchy into a graph-native computational object. Rather than treating Reactome as a set of sparse layer-to-layer constraints, PPI-Net treats hierarchy levels as node types and biological parent–child relationships as typed edges, enabling attention-based message passing across biological scales. This "conveyor belt" design preserves the interpretability of pathway-guided architectures while allowing the flexibility of graph neural networks to learn context-specific routes of information flow.

The results of multi-omics breast cancer classification suggest that the phenotype captured by PPI-Net is driven by the coordinated activation of stress-response, signaling and membrane excitability programs. At the molecular level, the model prioritizes canonical regulatory hubs, including the TP53–MDM2–AKT1 axis and HSP90-mediated proteostasis, reflecting dysregulation of apoptosis, cell cycle control and protein stability.[10,34,47,49] These signals propagate through the Reactome hierarchy to converge on higher-order pathways associated with cellular responses to stress and external stimuli, indicating that tumor cells operate in a state of sustained adaptive pressure.

In parallel, the model consistently identifies ion channel modules, including potassium and calcium channel families, which map onto neuronal system pathways at the pathway level. This suggests that cancer cells may co-opt membrane excitability and ion transport mechanisms to regulate intracellular signaling, proliferation and survival. Although ion channel dysregulation has been implicated in specific cancer contexts[51,54,57], its consistent emergence across multiple tumor types in our analysis points to a broader and potentially underappreciated role in tumor biology.

At the systems level, these programs are linked to metabolic adaptation and protein turnover pathways, indicating a coordinated network in which stress signaling, membrane dynamics and resource management collectively support tumor growth. This multi-scale convergence implies that cancer is not driven by isolated pathways, but by the integration of regulatory, metabolic and electrophysiological processes that enable cellular adaptation to a dynamic microenvironment.

From a methodological perspective, PPI-Net demonstrates that integrating molecular interaction networks with hierarchical pathway knowledge can improve both predictive performance and interpretability. Taking the PPI-Net approach improves classification of breast cancer patients using RNA-seq by 6.7% balanced accuracy compared to PPI alone. By structuring the model around biologically meaningful relationships, the approach reduces the reliance on purely data-driven feature learning and instead constrains representations to reflect known biological organization. This enables direct tracing of model predictions from proteins to genes and pathways, providing a mechanistic framework for hypothesis generation. The incorporation of multi-omics data further refines these representations, suggesting that epigenetic information can contextualize transcriptional signals within disease-relevant regulatory programs.

Additionally, our methodology demonstrates the strength of multi-headed decision making from multiple molecular levels. Compared to decision making from the single, final output layer of Reactome, our multi-headed approach had a relative improved balanced accuracy of 12.3%. This implies that both early and late scales for biological pathways are relevant in influencing disease state, and should be taken into account when developing future multi-scale disease models.

Despite these strengths, several limitations should be considered. First, the model relies on curated interaction and pathway databases and is therefore dependent on the completeness and accuracy of these resources. Incomplete or biased annotations may influence both predictive performance and interpretation. Second, the Reactome hierarchy is incorporated as a fixed structure, which may not fully capture context-specific pathway rewiring in different tumor types or microenvironments. Third, although the attribution framework provides biologically interpretable signals, it remains computational and requires experimental validation to confirm causal mechanisms.

Future work could extend this framework by incorporating additional biological priors, such as gene regulatory networks or spatially resolved interactions, as well as by enabling dynamic or learnable pathway structures. Integration with clinical outcomes, such as survival or treatment response, may further enhance the translational relevance of the model. In addition, the consistent identification of ion channel–associated pathways highlights a potential area for further investigation, both computationally and experimentally, to determine their role as therapeutic targets across cancer types.

In summary, PPI-Net provides a generalizable framework for modelling disease across biological scales, linking molecular interactions to functional processes and system-level phenotypes. By combining predictive performance with mechanistic interpretability, this approach offers a pathway toward more transparent and biologically grounded machine learning models in cancer research.

**Funding**:

This work was supported by several grants, including the AIDA project, funded by UK Research and Innovation (Grant No. 10058099) and the European Union (Grant No. 101095359); an Imperial College President's Scholarship; and the Cancer Research UK Early Detection and Diagnosis Programme scheme (Grant Ref. EDDPGM-May21\100007); Pancreatic Cancer UK, grant number BT2022_Hanna.

**Code Availability Statement**:

Code is available upon request by contacting corresponding author and will be made fully available to public bitbucket repository upon peer-reviewed publication.

**Author contributions**:

K.H. was responsible for methodological development, simulation design and experiments, evaluation of results, and manuscript preparation. G.G, D.V., and I.L. contributed to the conceptualization, methodology development, and manuscript review. K.V. supervised the methodological development, simulations, and evaluations, and was responsible for securing funding and overall project management. All authors contributed to the writing and editing of the manuscript and approved the final version.

| Biological Program | Key Genes / Components | Functional Role in Cancer | Evidence from Model | Key References |
|---|---|---|---|---|
| **Proteostasis / Heat Shock Response** | HSP90AA1, HSP90AB1, HSF1 | Stabilizes oncogenic proteins (TP53, AKT, ER, RTKs); supports survival under proteotoxic stress | Strong pathway flow into stress-response and external stimuli programs | 34–37, 38, 39 |
| **Ubiquitin–Proteasome System** | UBB, RPS27A | Protein degradation, stress adaptation, regulation of HSP90 clients, proliferation control | High downstream contribution to stress and external stimuli pathways | 40, 41, 42, 43, 58, 59 |
| **TP53–MDM2–AKT Axis** | TP53, MDM2, AKT1 | Controls apoptosis, proliferation, DNA damage response; drives tumor progression when dysregulated | Central multi-scale module linking PPI and pathway hierarchy | 44–49 |
| **Replication / Cell Cycle Stress** | CDK2 (via pathways), pre-replicative complex | Enables proliferation under stress; linked to genomic instability | Enriched within stress-response hierarchy | 44–47 (contextual), broader replication stress literature |
| **Oxidative Stress & Senescence (SASP)** | Oxidative stress pathways, SASP-related signaling | Drives inflammation, microenvironment remodeling, tumor progression | Contributes to cellular response to stress programs | (contextual, supported by broader senescence literature) |
| **Ion Channel / Bioelectric Signaling** | KCNK (1,3,9,10), KCNJ (3,11), KCNN (2,4) | Regulates membrane potential, proliferation, migration, metabolism | Strong contribution to neuronal system pathways | 50–57 |
| **KCNK (Two-pore potassium channels)** | KCNK1, KCNK3, KCNK9, KCNK10 | Proliferation, invasion, hypoxia resistance, prognosis | Enriched across multiple pathways | 50, 51, 52 |
| **KCNJ (Inward rectifier channels)** | KCNJ3, KCNJ11 | Prognostic biomarkers; hormone-linked signaling | Highlighted in pathway flow | 53, 54 |
| **Calcium-activated K+ channels** | KCNN2, KCNN4 | Links calcium signaling to proliferation, apoptosis, metabolism | Connects calcium and potassium signaling modules | 55–57 |

| Biological Program | Key Genes / Components | Functional Role in Cancer | Evidence from Model | Key References |
|---|---|---|---|---|
| **Metabolic Adaptation / Nutrient Processing** | Digestion & absorption pathways | Supports tumor growth via nutrient acquisition | Strong pathway contribution linked to proteasome activity | 58 |
| **Cytokine / Interleukin Signaling** | Interleukin pathways | Inflammation, immune modulation, tumor microenvironment | Present in stress-response convergence | (implicit in SASP / stress literature) |
| **Reproductive / Developmental Programs** | Developmental pathways | Dedifferentiation, cancer stem cell-like states | Identified in pathway hierarchy | 60 |
| **Vesicle-mediated Transport** | Trafficking pathways | Resource acquisition, membrane dynamics, signaling | Secondary but consistent program | (general cancer trafficking literature) |